# Title:

Programmable metachronal motion of densely packed magnetic artificial cilia

# Authors:

Tongsheng Wang[1,2], Tanveer ul Islam[1,2], Erik Steur[1,2], Tess Homan[1], Ishu Aggarwal[3], Patrick R. Onck[3], Jaap den Toonder [1,2], and Ye Wang*[1,2]

# Affiliations

[1] Department of Mechanical Engineering, Eindhoven University of Technology, 5600 MB, Eindhoven, The Netherlands
[2] Institute for Complex Molecular Systems, Eindhoven University of Technology, 5600 MB, Eindhoven, The Netherlands
[3] Zernike Institute for Advanced Materials, University of Groningen, 9747 AG, Groningen, The Netherlands
*E-mail: y.wang2@tue.nl

# Abstract

Despite recent advances in artificial cilia technologies, the application of metachrony, which is the collective wavelike motion by cilia moving out-of-phase, has been severely hampered by difficulties in controlling densely packed artificial cilia at micrometer length scales. Moreover, there has been no direct experimental proof yet that a metachronal wave in combination with fully reciprocal ciliary motion can generate significant microfluidic flow on a micrometer scale as theoretically predicted. In this study, using an in-house developed precise micro-molding technique, we have fabricated densely packed magnetic artificial cilia that can generate well-controlled metachronal waves. We studied the effect of pure metachrony on fluid flow by excluding all symmetry-breaking ciliary features. Experimental and simulation results prove that net fluid transport can be generated by metachronal motion alone, and the effectiveness is strongly dependent on cilia spacing. This technique not only offers a biomimetic experimental platform to better understand the mechanisms underlying metachrony, it also opens new pathways towards advanced industrial applications.

# Teaser

Densely packed artificial cilia performing metachronal motion are able to generate significant fluid flow.

# MAIN TEXT

# Introduction

Cilia exist ubiquitously in nature and play key roles in many biological functions, such as locomotion, hearing, feeding, immune responses and reproduction[1]. Inspired by nature, various types of artificial cilia have been developed[2] in the past two decades as sensors or actuators that can respond to or be driven by external stimuli, such as electrostatic fields[3], magnetic fields[4-12], light[13], pneumatic pressure[14-17], sound[18] and pH[19]. Most of them are used as *in-situ* pumps and mixers within microfluidic devices, and they most commonly use magnetic actuation due to the advantage of untethered power transfer and the compatibility with biological materials. The study of magnetic artificial cilia started with synchronous, non-reciprocal motion, which has been shown to be effective in producing fluid flow in a microfluidic environment [7, 9, 11, 12, 17, 20-22].

In recent years, there has been an increased focus in studying metachronal motion, i.e. the collective wavelike motion resulting from neighboring cilia moving slightly out-of-phase. Researchers have studied the origin and the functional effects of metachronal motion[23-25] by modeling or mimicking this signature behavior using artificial cilia, with the intention of eventually leveraging the insights towards practical applications. For magnetic artificial cilia, metachronal motion can be generated either by applying a spatially non-uniform magnetic field to an array of identical cilia[23], or by engineering different responses of each individual cilium to the same uniform magnetic field[24, 26-29]. For the first approach, the challenge lies in the generation of sufficiently large local variations in the magnetic field. Such fields can only be generated with an array of carefully arranged small magnets[23], but the field strength quickly diminishes with increasing distance to the array, making it impractical for real applications. The second approach, on the other hand, puts a challenge on the fabrication process, as the neighboring cilia need to have sufficiently different properties in order to generate regular and tunable phase differences. As a result, those artificial cilia are mostly fabricated at length scales of millimeters or larger[29-31], restraining these technologies to down-scale the dimensions as is more and more required in the miniaturization of microfluidic devices[2]. Another, perhaps more fundamental issue with the current approaches, is that the spacing between the artificial cilia are about the same as their length[23, 24], while natural cilia are much more closely packed[25]. This is not coincidental: when the magnetic cilia are too close to each other, the dipole-dipole interactions between them will interfere with their motion, causing the formation of ciliary pairs or bundles, after which their motion can no longer be controlled[4]. This is not a trivial problem, because interciliary spacing is critical to the mechanism of metachronal motion and reducing spacing can strongly increase fluid transport, as shown earlier by theoretical and numerical studies[32-38]. To overcome these drawbacks, we developed a well-controlled experimental set-up consisting of μm-scale magnetic artificial cilia at unprecedented interciliary distances.

In this work, we have fabricated densely packed magnetic artificial cilia with programmable metachronal motion, using a precise micro molding method. The magnetic dipole interaction between cilia was minimized by making only the base part magnetic, allowing us to fully control their motion while having them closely spaced. The base of every cilium follows the shape of the mold with specifically designed inclination angles, which determine the phases of their motion during the actuation. The molds were made from fused silica with a femtosecond laser assisted etching (FLAE) process, enabling precise control over the geometries. Then, artificial cilia were made by transfer molding two layers of precursor materials with controlled thicknesses into the glass mold: a magnetic layer forming the base part of the cilia, and a nonmagnetic layer forming the top part. The cilia were actuated by an in-house magnetic actuator set-up and the motion and flow generated by various cilia arrays were analyzed using highspeed imaging and particle tracking, and the results were compared with and supported by numerical simulations that fully couple the effects of elasticity, magnetism and fluid dynamics.

## Results

### Fabrication and actuation of the densely packed artificial cilia

The cilia fabrication process involves a femtosecond laser-assisted etching step[39] and a subsequent two-layer molding step. As shown schematically in Fig. 1A(1-3), after laser machining a pre-designed path into a fused silica slide using a focused femtosecond laser (FEMTOprint f200), the slide was put in an 85 °C ultrasonic bath of a 45 wt% potassium hydroxide solution to remove the modified part. Using this method, all the geometrical parameters can be adjusted and realized with micrometer accuracy, as shown in Fig. 1(B). Two molds are shown here, which were used to fabricate cilia that can generate opposite metachronal motion. For the so-called divergent array shown in Fig. 1B (1), the base inclination angles $\theta$ change from 135° to 45° in steps of 10°. Similarly, in Fig. 1B (2), the inclination angles change in the opposite way, forming a so-called convergent array.

A two-layer transfer molding technique was developed to fabricate partially magnetized cilia. Fig. 1A (4-6) shows the schematics of the molding and demolding process. Figure 1C shows two different arrays of cilia that contain the same individual cilium but in different arrangements, corresponding to the molds shown in Figure 1B. The black part of the cilia is magnetic and the transparent part is nonmagnetic. These cilia have a width $c_w$ = 24 μm, length from tip to substrate $c_l$ = 500 μm, depth $c_d$ = 200 μm, and tip-to-tip pitch $\delta$ = 200 μm; the length of the inclined part $l_m$ is 100 μm for all the cilia.

The demolded cilia patch was then integrated into a microfluidic chip containing a recirculation channel, as illustrated in Fig. 1D. The channel was also made with FLAE using a 1 mm thick fused silica glass slide. After integrating the cilia patch, the channel side was capped off with a flat glass slide. On the back of the channel side, it has two openings for filling liquid with tracer particles, which can be sealed with tape afterwards (see SI Fig. S-7). The assembled chip was then placed on the chip holder, where the cilia patch was in the central area of a Halbach array (Fig. 1E). The Halbach array consists of 16 neodymium magnets of $10 \times 10 \times 40$ mm, arranged in a specific way to generate a uniform and unidirectional magnetic field[40]. Figure 1F shows a simulation result of the magnetic field, where the field strength in the central area is 0.22 T, which is validated by measurement using a gaussmeter. The Halbach array is rotated using a DC motor with a timing belt, as shown in Fig. 1G, with the speed up to 6000 rpm, allowing us to apply a highly controlled rotating uniform field to the cilia at different frequencies.

### Individual and collective motion

There are two main mechanisms for flow generation by artificial cilia when they are moving synchronously (no metachrony), namely the spatial asymmetry of the cilia motion and the effect of inertia in the case of temporal asymmetrical motion. In both cases, the motion can be divided into an effective stroke, where more fluid is driven in one direction, and a recovery stroke, where less fluid is driven back. For fully magnetic cilia in the previous studies[9, 11, 15, 20, 23, 24, 29, 41], spatial asymmetry was the main mechanism in net flow generation, while inertia has also been found to induce or enhance the net flow in some cases, when during the effective stroke the local Reynolds number ($Re$) at the cilia tip reaches a value much larger than 1[3, 24].

Interestingly, in all the experimental studies on flow generation by cilia undergoing metachronal waves, these asymmetries are always present when a significant net flow is observed. This is partially by design, aimed at further enhancing the net flow generation, especially for millimeter sized cilia[15]. However, in the case of smaller cilia with dimensions of hundreds of micrometers or smaller, these effects are intrinsic to their dynamics and cannot be turned off. This is because the motion is always slower when the cilia are following the magnetic field, and faster when they bounces back elastically, inducing temporal asymmetry as well as spatial asymmetry, due to different levels of drag induced bending between the two strokes.

In this work, the motion of cilia are both spatially and temporally symmetrical and there is no effective or recovery stroke, allowing us to study the effect of pure metachrony. This is achieved by the two segment design, in which the magnetic root controls the phase of the motion, and the nonmagnetic part provides enough damping to eliminate both asymmetries.

Fig. 2A shows a superimposed image of a cilium at three timepoints in one beating cycle under a clockwise rotating magnetic field. During the magnetic stroke, the torque bends the cilia root to the right and the rest of the non-magnetic cilia body follows this motion. As the magnetic field rotates, the elastic torque builds up until it exceeds the magnetic torque, and the cilium bends back, performing the elastic stroke. Fig. 2B shows the tip trajectories of all cilia, with both experimental and simulation results. It can be seen that for each cilium, the tip trajectories during the magnetic and elastic strokes collapse onto the same line, without an enclosed area, so there is no spatial asymmetry in motion.

Fig. 2F and G show the *x*-position of the cilia for the two types of arrays, respectively, with results from both experiments and simulations. It can be clearly seen that the two arrays generate metachronal waves in opposite directions, in which the movement profiles are similar but shift incrementally with regular intervals. Moreover, the profiles are roughly sinusoidal without sudden movement, as a result of the damping of the elastic stroke by the non-magnetic part. Fig. 2H and I show the local *Re* at the cilia tip, which is defined as $Re = \rho u c_l / \mu$, where $\rho$ is the density of fluid, $u$ is cilia tip speed, $c_l$ is cilia length, and $\mu$ is the dynamic viscosity. It can be seen that *Re* is small at all times, meaning that the inertial effects are thus negligible. In this way, the effect of temporal asymmetry and inertia are also eliminated.

Note that the beating amplitude is sensitive to the magnetic filling height, which is defined by the distance from the top of the magnetic part to the substrate. The variation in the height of the magnetic part for each cilium in the arrays is very small, about ± 10 μm, as shown in Fig. 2C. However, due to the difference in the inclination angles, the cilia on the far ends of the arrays have longer magnetic parts, which results in higher beating amplitudes for these cilia, as shown in Fig. 2D and E. Note that these variations do not induce significant shape anisotropy or affect phase differences of the metachronal motion. As the actuation frequency increases, the increased hydrodynamic drag reduces the amplitude of the motion, as shown also in Fig. 2D and E.

## Characterization of local vortices and global net flow

By eliminating motion asymmetry and inertial effects, any net flow generation in our system can only be attributed to metachronal waves. We performed detailed analysis of both local and global flow. Fig. 3A and B show the time-dependent flow fields containing what can be described as walking vortices, for the divergent and the convergent array, respectively. As shown both by the experimental results (left column) and the simulation results (right column) at an actuation frequency of 2 Hz, the streamlines form two counterrotating vortices above the cilia arrays, and these vortices travel in the same direction as the metachronal wave. Note that for both the divergent and the convergent cilia arrays, the magnetic strokes are all toward the right and the elastic strokes are toward the left because in both cases the cilia are subjected to a clockwise magnetic field. However, the propagation direction of the metachronal wave follows the magnetic stroke for the divergent array but the elastic stroke for the convergent array, which means that the vortex travel direction is determined by the wave propagation direction, rather than the individual cilia motion. This is important because this means that if a global net flow is generated purely by metachrony, one would expect that these two arrays will generate opposite net flow with the same amplitude, which is indeed the case as shown below.

Fig. 4B shows the tracer particle trajectories in the recirculation channel (ROI 1 in Fig 4A) for the divergent cilia array actuated at 100 Hz. Fig. 4C shows the velocities derived from Fig. 4B, which fits well to the theoretically derived flow profile[42]. Fig. 4D and E show particle trajectories above the two arrays. In each of the panels, four cycles of particle movement at different *y*-levels are shown. It can be seen that the traces of particles are non-reciprocal, and they show different net displacement on different height levels. The time averaged particle velocities in the *x*-direction with respect to the vertical distance above cilia tips are shown in Fig. 4F, where the largest velocity appears near the cilia tips with the same direction as the overall net flow, and the velocities decrease and reverse as the distance increases. Note that the flow profiles are mirrored for the two different arrays, despite the fact that all cilia in both cases have the same magnetic and elastic stroke directions, confirming that the flow is entirely due to metachrony.

To study the time-integrated flow profile, we performed longer duration particle tracking using stroboscopic imaging with the framerate matching the cilia beating frequency. The results are shown in Fig. 4G and H; the corresponding movies and recording parameters can be found in SI. Different from the dynamic walking vortices from the time-dependent instantaneous flow fields, stroboscopic imaging reveals stable, larger scale flow patterns, which are similar for the two arrays, but in opposite directions. Note that the flow rate achieved by these cilia are comparable to that achieved in the literature with asymmetrical cilia motions. These nearly mirrored flow patterns and the significant net flows generated by reversing the wave direction prove that metachronal wave alone can generate effective fluid pumping, without the need for individual cilium performing any asymmetrical motion.

## Effect of actuation frequency and cilia pitch on flow generation

The flow structures at higher frequencies visualized using stroboscopic imaging are shown in Fig. 5A and B. The particle trajectories show the local net movement over 3 seconds. Due to the increased fluid drag reducing the amplitude of cilia motion as shown in Fig. 5C, the amount of fluid transported per beating cycle is reduced with increasing beating frequency as shown in Fig. 5D. However, the increase in the beating frequency compensates for the loss, resulting in a nonlinear increase of the volume flow rate at higher frequencies measured in the recirculation channel, as shown in Fig. 5E. Fig. 5F shows the peak $Re$ at different frequencies. It can be seen that the $Re$ exceeds 1 at higher frequencies, reaching about 5 at 100 Hz, which is still relatively low. Indeed, if inertia is significant, the flow generated by the two arrays would not be mirrored anymore at high frequencies, since they share the same individual cilia motions. As shown in Fig. 5E, with the two arrays showing mirrored flow rates, it can be concluded that inertia can be ignored for net flow generation even at high frequencies.

To investigate the effect of cilia spacing, we performed 2D numerical simulations using COMSOL Multiphysics®. The results are shown in Fig. 6. A 2D model is chosen for computational efficiency and it can sufficiently represent the flap-like shaped cilia, which perform 2D motion, and generate mostly 2D flow patterns. In the model, the cross-sectional geometries of the cilia and the magnetic filling height are the same as in the experiments shown in Fig. 2E. Validation was first performed using experimental results from both convergent and divergent arrays with a cilia pitch of 200 μm to ensure the reliability of the simulation results. These comparisons between simulation and experiments have already be shown in Fig. 2 and 3.

It can be seen from Fig. 6A that only when below 200 μm, the spacing of the cilia starts to significantly affect their beating amplitude. This means that the hydrodynamic interaction forces between cilia are relatively small compared to the magnetic and elastic forces, and a significant change in the generated net flow with respect to the spacing, which will be shown below, cannot be attributed to the changes in the individual cilium motion.

Fig. 6B and C show the particle tracking results from the simulation of divergent cilia arrays with 200 and 500 μm spacings, respectively. It can be seen that Fig. 6B matches the experimental result shown in Fig. 4G, indicating the validity of the simulation. Comparing Fig 6B and C, it can be seen that the vortex structure diminishes when the spacing increases, and the predicted volume flow rate decreases as a result, as shown in Fig. 6D. Note that in earlier simulation studies, similar effects of cilia spacing on flow generation have been found using a different numerical framework[36]. Our current results experimentally demonstrate and numerically confirm these earlier studies that fluid flow can indeed be generated by reciprocal cilia subject to metachronal waves. These results further highlight not only the exciting opportunity of creating densely packed cilia for enhanced fluid flow, they also can be used as an experimental biomimetic platform for studying the fundamental mechanisms underlying metachronal motion.

## Discussion

In this study, we have created densely packed magnetic artificial cilia using a femtosecond laser-assisted etching process combined with a double-layer micro molding technique. In order to study the effect of pure metachronal waves on flow generation, we have engineered our system to eliminate the effects of spatial and temporal asymmetry in individual cilia motion. This has been achieved by designing the cilia with a magnetic base, which controls the phase of their motion, and a nonmagnetic body, which introduces high hydrodynamic damping, resulting in reciprocal individual cilia motion. We have performed detailed studies of the cilia motion, flow patterns and net flow generation for opposite metachronal waves under different actuation frequencies, using both experiments and simulations. We also performed further analysis of the effect of cilia spacing on flow generation with experimentally validated numerical simulation.

The results provide a definitive proof that metachronal motion alone can generate a significant net fluid flow, without nonreciprocal individual cilia motion or any inertial effects. A strong influence of the spacing between cilia on the flow generation is found, where the flow only becomes significant when the cilia are close together. This also explains why such proof was not shown in the literature so far, because almost all metachronal ciliary systems created in the past, including ours, were sparsely distributed, with the gaps between the cilia similar to or larger than their lengths. It shows the importance of using such a densely packed cilia system as an experimental platform for studying important physiological functions of metachronal waves, such as particle and viscoelastic fluid transportation. In addition, the magnitude of the flow generated in this work is comparable to other artificial cilia that rely on nonreciprocal motions[2], which brings a fundamental shift to the design principles of artificial cilia based *in-situ* flow generation systems for microfluidic applications.

Moreover, the fabrication method developed in this work enables unprecedented level of control over a combination of geometry and distribution of materials of different properties in 3D on a micrometer scale, which opens up the possibilities for creating micro-actuators and sensors with hitherto unrealizable degrees of freedom in designing and controlling their motion. This method can also be applied for fabrication of devices made from other types of polymeric materials, such as light and temperature responsive ones. As a replica molding technique, which is beneficial for production scale-up, this approach has the potential to bring profound impact to Lab-on-a-Chip as well as other fields of application, such as dynamic surface topographies[43-46] and microrobotics[47, 48].

# Materials and methods

## The fabrication process of cilia mold and channel

A femtosecond laser with a wavelength of 1030 nm, repetition rate of 1000 kHz, and 230 nJ pulse energy was used to induce local microstructure changes in a 76×26×1 mm fused silica glass slide along the designed paths. The laser system is integrated into FEMTOprint f200 aHead. The laser is focused through a Thorlabs 20x lens, resulting in an effective voxel with an ellipsoidal shape, which has a 3 µm short axis in the x-y plane and a 24 µm long axis in the z direction.

The laser tool path is generated in Alphacam Ultimate Mill software. Using Rough/Finish, Pocketing and 3D machining tools in the software and a well-designed tool path strategy, the laser writing of 3D structures is possible. After laser machining, the glass slides are put in a 45 wt% KOH bath for the etching process. The machined part has a faster etching speed, roughly 130 µm/h, than the unmachined part which is 0.7 µm/h, resulting in removal of the laser written structures.

Both the cilia mold and the microfluidic channels were made with FLAE. After etching, the channel was ready for integration and no further processing was performed. The mold was cleaned in an ultrasonic bath with DI water and dried for surface silanization with a drop of trichloro(1H,1H,2H,2H-perfluorooctyl)silane in a vacuum desiccator overnight.

## Transfer molding process

The molding was performed using a Specac hydraulic presses with a heating unit, as shown in SI 2. A premade thin layer of poly(Styrene-block-Isobutylene-block-Styrene) or SIBS (Kaneka Corporation), was first placed on the mold, which was then covered by another premade thin layer of magnetic SIBS containing 70 wt% superparamagnetic carbonyl-iron powder from Sigma-aldrich company. The mixing was performed using a twinscrew extruder mixer, shown in SI 2.1. A PTFE sheet was placed to separate the top layer and the compression plate to prevent adhesion. The thicknesses of the pure and magnetic SIBS layers were fine-tuned to achieve a desirable magnetic filling height (SI 2.2). In this particular case, they were 70 µm and 280 µm, respectively. The mold and the two layers of SIBS were first heated to 130 °C, then a force of 4 kN was applied to press the layers into the mold. After 15 minutes, the temperature was decreased to 90 °C, at a rate of 2.5 ~ 3 °C/min. After that, the pressure was released and the cilia patch was manually demolded while the mold was submerged in isopropanol.

## Visualization of the flow field

The flow field was observed through an Olympus microscope, and the videos were captured by a camera connected to it. The cilia motion and the vortex structures were measured at ROI 2, as indicated in Fig. 1D, and the fully developed flow was measured at ROI 1. SI4 shows the information of lens, cameras, and capture fps. We used a high frame rate to record the beating of cilia and walking vortices, and a low frame rate for the stroboscopic view of the vortex. Tracing particles (5 µm carboxyl-functionalized polystyrene particles, microParticles GmbH) were used to visualize the flow field. The recorded videos were analyzed in Fiji software with the TrackMate[49, 50] plugin. The tracking results were then transferred to MATLAB for further analysis.

## COMSOL simulation model

To help the analysis of cilia motion and the flow structures, we developed a 2D COMSOL® simulation model, fully coupling magnetism, solid mechanics and fluid mechanics. The governing equations are as follows:

$$\nabla \cdot \mathbf{B} = 0 \tag{1}$$

$$\mathbf{H} = -\nabla V_m + \mathbf{H_b} \tag{2}$$

$$\mathbf{B} = \mu_0 \mu_r \mathbf{H} \tag{3}$$

$$\rho_s \frac{\partial^2 \mathbf{u_s}}{\partial t^2} = \nabla \cdot \mathbf{S} \tag{4}$$

$$\nabla \cdot \mathbf{v_f} = 0 \tag{5}$$

$$\frac{\partial \mathbf{v_f}}{\partial t} + (\mathbf{v}_f \cdot \nabla)\mathbf{v}_f = \mathbf{f} - \frac{1}{\rho} \nabla p + \upsilon \nabla^2 \mathbf{v_f} \tag{6}$$

$$\boldsymbol{\sigma} \cdot \mathbf{n} = \mathbf{f} \cdot \mathbf{n} \tag{7}$$

$$\mathbf{v_f} = \frac{\partial \mathbf{u_s}}{\partial t} \tag{8}$$

Equations (1, 2) are utilized in magnetostatics to determine the magnetic field in the absence of a current. $\mathbf{H}$ is the magnetic field strength, $V_m$ is the magnetic scalar potential and $\mathbf{H_b}$ is the background magnetic field strength. In the simulations, $H_{bx} = H_b sin(2\pi ft), H_{by} = H_b cos(2\pi ft)$, where $f$ is the actuation frequency and $t$ is the time. A zero scalar potential is assigned to an arbitrary boundary. The magnetic flux density, $\mathbf{B}$, is measured in Tesla and is related to $\mathbf{H}$ by equation (3), where $\mu_0 = 4\pi \times 10^{-7}$ H/m. For magnetic SIBS, $\mu_r = 1.32$, which is fitted from comparing simulation to the experimental results, whereas it is 1 for pure SIBS and water. Equation (4) represents the solid mechanics equation, where the solid density $\rho_s = 970 \text{ kg/m}^3$, $\mathbf{u_s}$ is the solid displacement field, and $\mathbf{S}$ represents the Cauchy stress tensor. The Young's modulus $E = 0.7$ MPa, and the Poisson ratio $\gamma = 0.49$ are given by the product company (Kaneka Corporation). In the simulation, equations (5, 6) describe the fluid continuity and momentum, where $\mathbf{v_f}$ is the fluid velocity, $\rho$ is the fluid density, $p$ is the fluid pressure and $\upsilon$ is the fluid kinematic viscosity. Equations (7, 8) are the arbitrary Lagrangian-Eulerian (ALE) method, where $\boldsymbol{\sigma}$ is the solid stress and $\mathbf{f}$ is the fluid force on the boundary of solid. The no-slip boundary condition is set on the cilia surfaces and on the wall of the channel, whereas an open boundary condition is applied at the inlet and outlet. The magnetic force acting on the cilia is determined using the Maxwell surface stress tensor. The equations are fully coupled and the MUMPS solver is used. The time discretization is Backward Differentiation Formula (BDF) with maximum order of 5.

## Author contributions

Conceptualization: TW, YW; Conducting experiments and simulations: TW; Designing the experiment setup: TuI; Collecting data and analyzing: TW; Writing and editing: TW, YW, TuI, ES, TH, IA, PRO, JdT; Supervision and financial support: JdT, YW

## Conflicts of interest

The authors declare that they have no competing interests.

## Acknowledgements

The research leading to these results has received funding from the European Research Council (ERC) under the European Union's Horizon 2020 research and innovation program under grant agreement no. 833214 and no. 953234

All data needed to evaluate the conclusions in the paper are present in the paper and the Supplementary Materials.

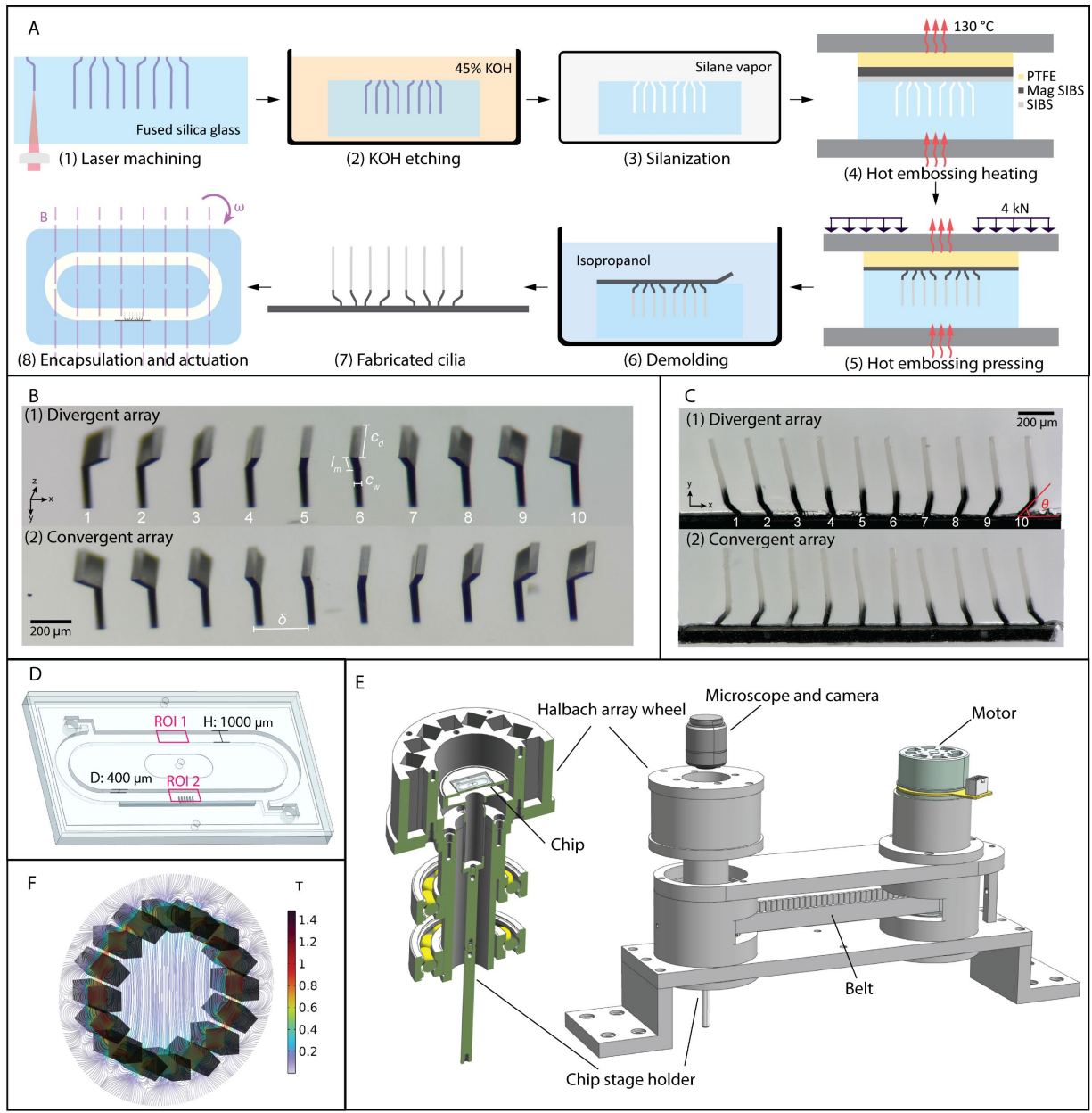

Figure 1: Fabrication of densely packed magnetic artificial cilia and the experimental setup. (A) Cilia fabrication process. (1) Local modification of fused silica slide using femtosecond laser writing (FEMTOprint f200). (2) Etching in 45 WT% potassium hydroxide with ultrasonication at 85 °C, to obtain a mold. (3) Overnight mold silanization with trichloro(1H,1H,2H,2H-perfluorooctyl)silane in a vacuum desiccator. (4) Mold filling by transfer molding at 130 °C with a double-layer sheet consisting of a layer of magnetic particle doped poly(Styrene-block-Isobutylene-block-Styrene), or magnetic SIBS, on top of a layer of pure SIBS. (5) Pressing the two layers into the mold under 4 kN. (6) Demolding in isopropanol. (7) Fabricated cilia patch. The black part is magnetic and the rest is non-magnetic. (8) Assembling the cilia patch in a chip and actuating by a rotating uniform magnetic field. (B) Picture of the molds after etching. Two types of molds with different configurations are machined. (1) is a divergent array and (2) is a convergent array. (C) Fabricated cilia with width $c_w = 24\ \mu m$, depth $c_d = 200\ \mu m$, and tip-to-substrate height of $500\ \mu m$. The inclined root angle $\theta$ changes from 135° to 45° from left to right in equal increments for the divergent array and from 45° to 135° for the convergent array. (D) Assembled chip containing a recirculating channel and a cilia patch. The height of the channel is 1000 µm and the depth is 400 µm. Two Regions of Interest are imaged. ROI 1 is for measuring fully developed net flow, and ROI 2 is for characterizing cilia movement and local flow patterns. (E) The experimental setup. The chip is placed on a stage located in the central area of the Halbach array wheel. The wheel is fixed on a shaft supported by two bearings and is driven by a motor with a connecting belt. (F) Simulation result of the magnetic field generated by a Halbach array. The magnetic flux density in the center area is 0.22 T, same as the measured value using a gaussmeter.

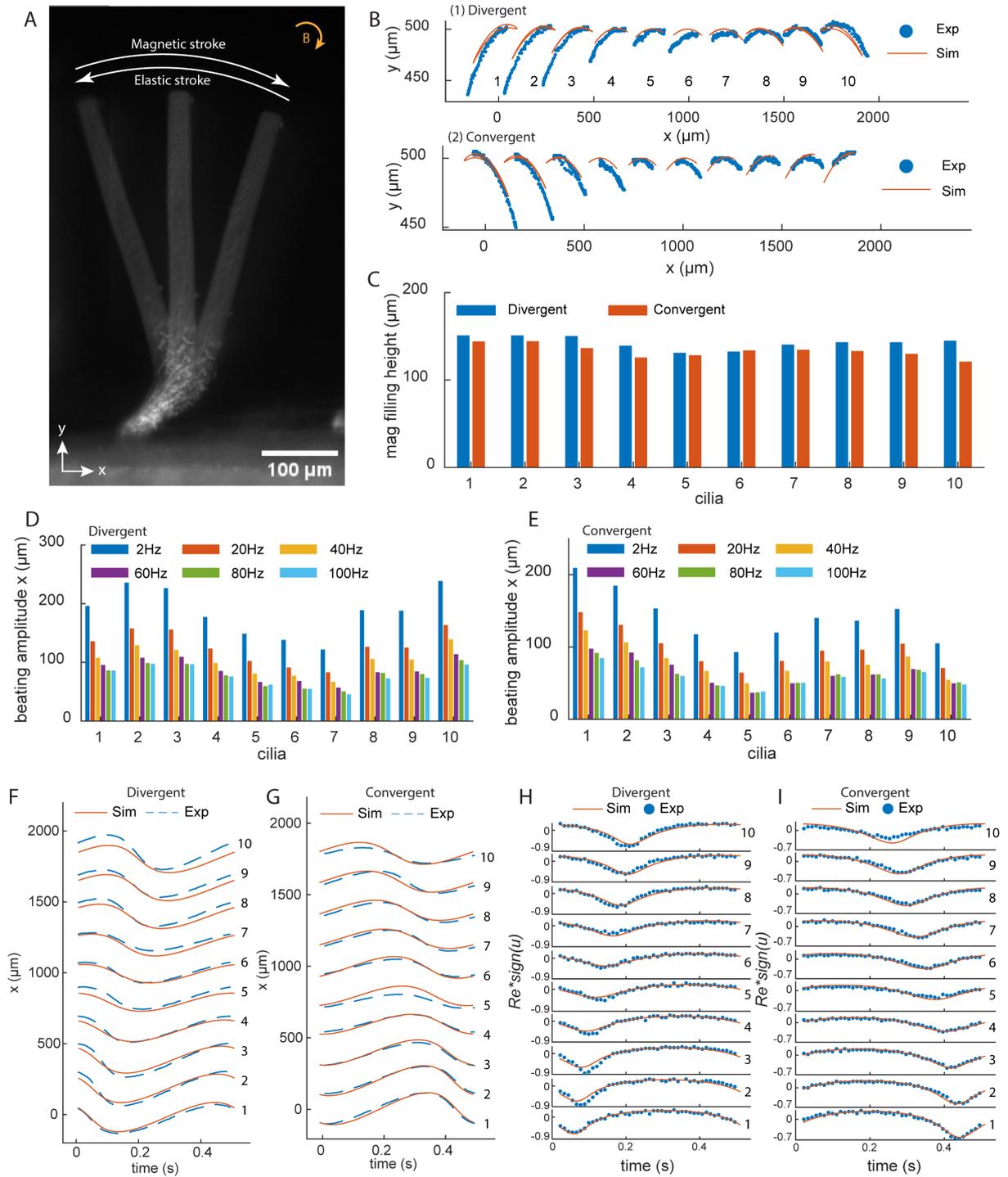

Figure 2: Cilia response to a clockwise rotating uniform magnetic field. (A) Superimposed images of a cilium from three timepoints in one beating cycle; the cilium has a base angle of 45° and is actuated at 2 Hz. The figure depicts the leftmost position, resting position and rightmost position. (B) The trajectories of the tips of the cilia. The blue dots are the measurement results from the experiment and the red solid lines are the simulation results obtained using COMSOL® Multipysics. (C) The measured height of the magnetic section of different cilia used in the experiments corresponding to the other panels in this figure. (D-E) The beating amplitude in the *x*-direction under different frequencies. (F-G) The experimental and simulation results of the cilia tip *x*-position under a cycle for all 10 cilia in the arrays. It shows phase differences between cilia and the wave propagation direction. (H-I) The experimental and simulation results of instantaneous cilia tip Reynolds number for all 10 cilia in the arrays. For better representation, the tip Reynolds number is multiplied by the direction sign of cilia tip velocity *u*.

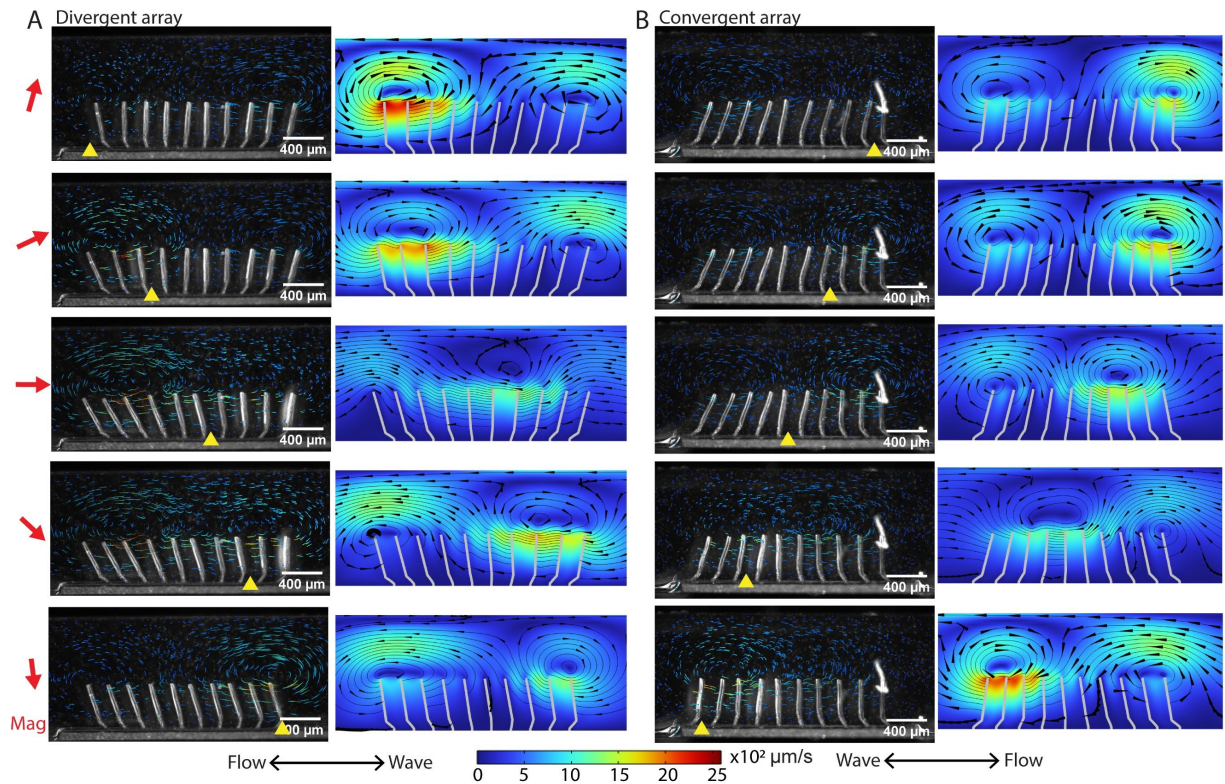

Figure 3: Highspeed imaging of metachronal motion and the resulting velocity field showing the 'walking vortices'. (A-B) The results of the experiment (left column) and simulation (right column) of the metachronal motion and the resulting vortices for the divergent array (A) and the convergent array (B), actuated at 2 Hz. The images are captured at a frame rate of 180. The red arrow on the left side of the figure is the magnetic field direction at the corresponding moments. The yellow triangle on the bottom of the image is the position where the cilia engage in the elastic stroke, when they can no longer follow the magnetic field and start moving in the opposite direction. In the experiments, tracer particles with a diameter of 5 μm are used to visualize the velocity field. The tail-like lines in the experimental images are the trajectories of the tracer particles during the past 30 frames or 0.17 seconds, and the color indicates the magnitude of velocity. In the simulation results, the streamlines show the instantaneous velocity field. The arrows are the velocity direction and the color is the magnitude. The corresponding videos can be found in the SI. These results show that divergent and convergent arrays generate opposite metachronal waves with the same magnetic field, and they generate similar patterned walking vortices in opposite directions.

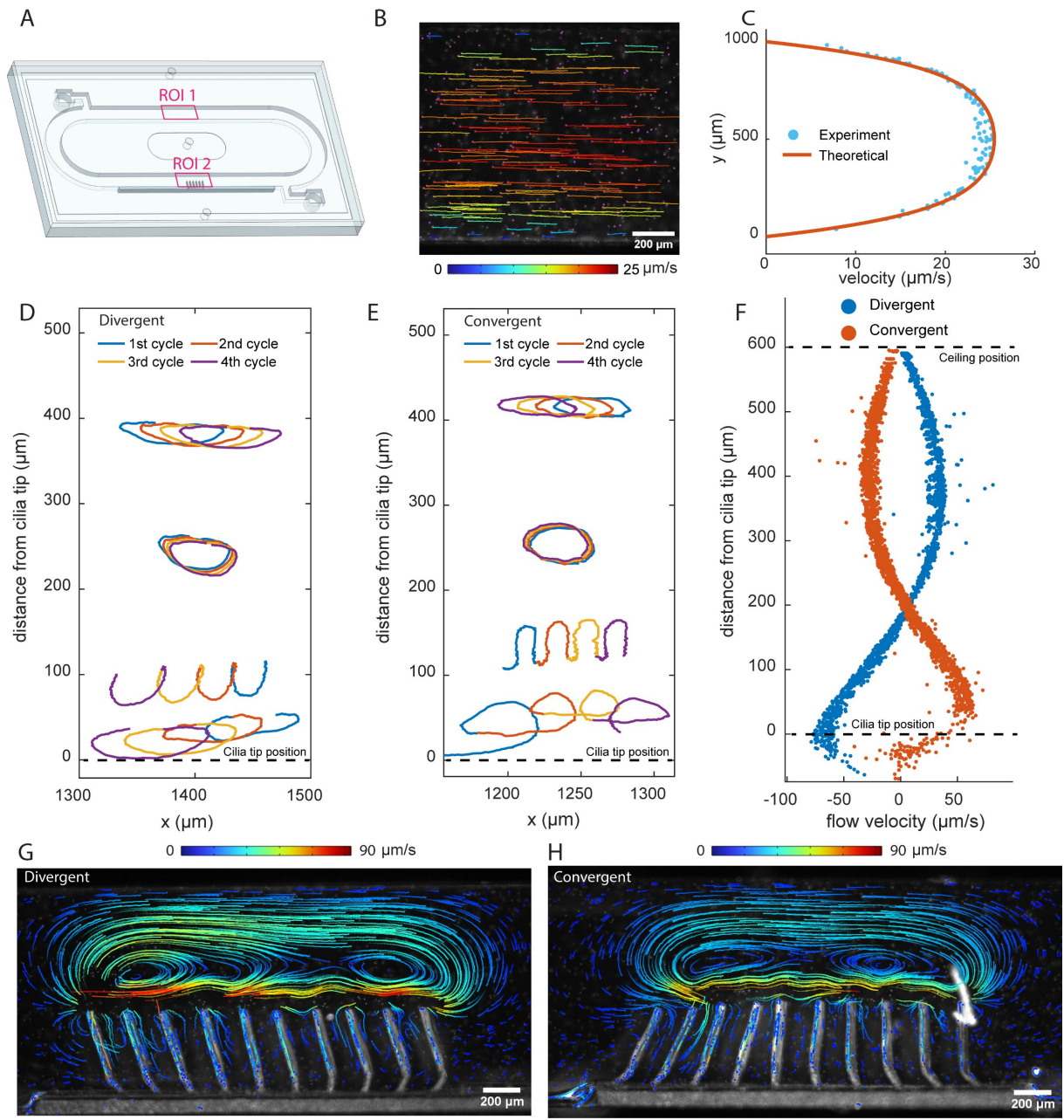

Figure 4: Net flow generation and particle tracing. (A) Schematic of the assembled chip containing a recirculating channel and a cilia patch. ROI 1 corresponds to (B-C), and ROI 2 corresponds to (D-H). (B) Particle trajectories of the fully developed flow with the divergent cilia array actuating at 100 Hz. (C) The velocity profile along the channel height using the divergent array at an actuation frequency of 100 Hz. (D-E)Tracer particle trajectories imaged at 2 Hz of beating frequency above cilia tips. The trajectories of four cycles represent typical trajectories at different distances above the cilia tip. (F) The net local flow velocities in x direction above the cilia tip derived from particle tracing. (G-H) The stable vortical streamlines obtained with stroboscopic imaging at a beating frequency of 2 Hz. They show opposite, almost mirrored vortex structures. Note that although the instantaneous velocity fields show repetitive moving vortices, as shown in Fig 3, long-term tracing of particles show static and stable streamlines, as shown here using stroboscopic imaging. The flow direction close to the cilia tip is the same as the overall net flow direction, which is to the left for the divergent array and to the right for the convergent array.

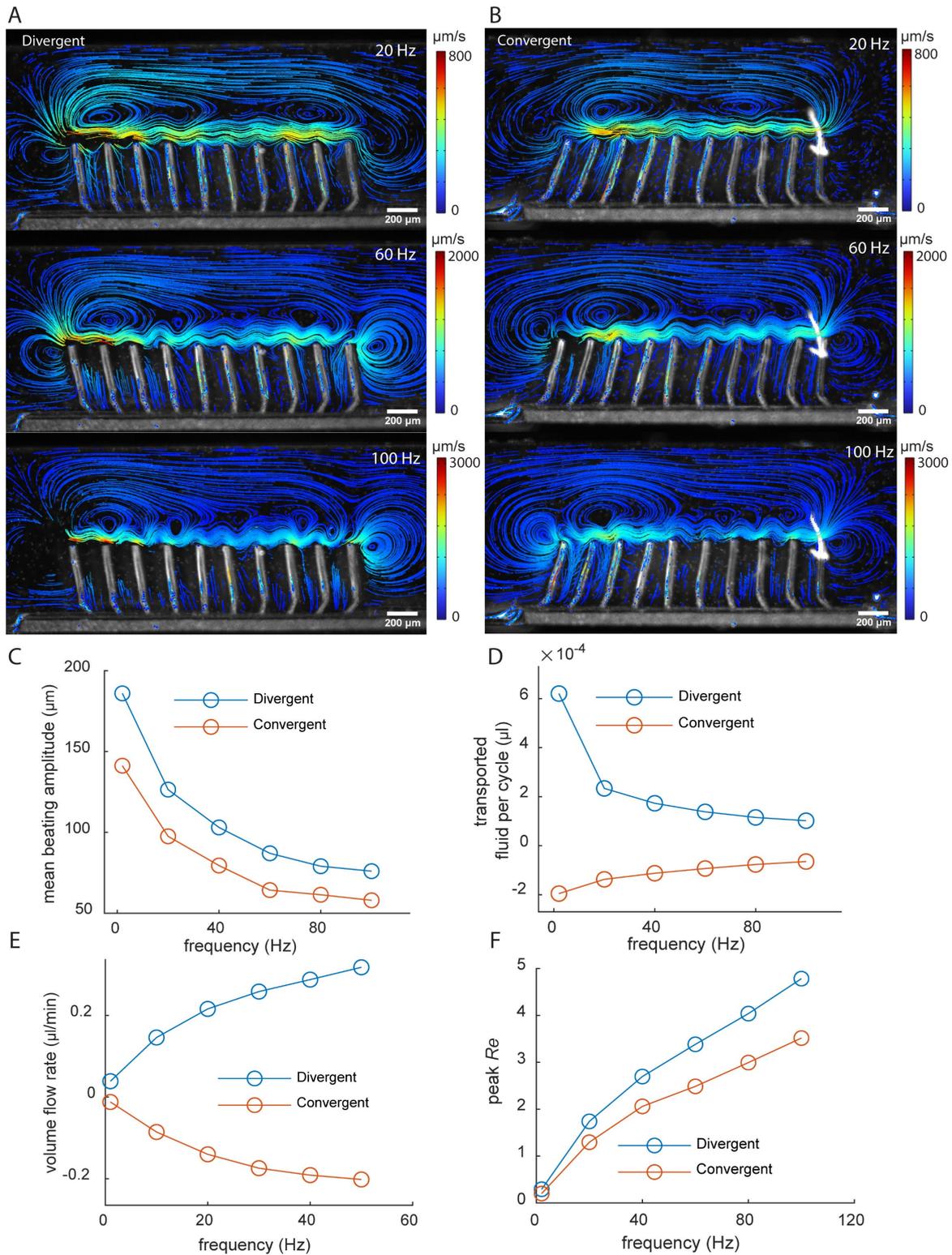

Figure 5: Particle streamlines and fluid transport at different frequencies. (A-B) The vortex patterns of the divergent array and the convergent array at different frequencies. The trajectories show the movement of the particles during 3 s. It can be seen that the shape of the vortices remains similar across frequencies, while the velocity increases. (C) The mean beating amplitude in the x direction of all cilia at different frequencies. (D) The transported fluid per beating cycle at different frequencies. (E) The volume flow rate at different frequencies. (F) The average peak $Re$ of all cilia at different frequencies. $Re = \rho u c_l / \mu$, where $\rho$ is the density of surrounding fluid, $u$ is cilia tip speed, $c_l$ is cilia length, and $\mu$ is the dynamic viscosity of the surrounding fluid.

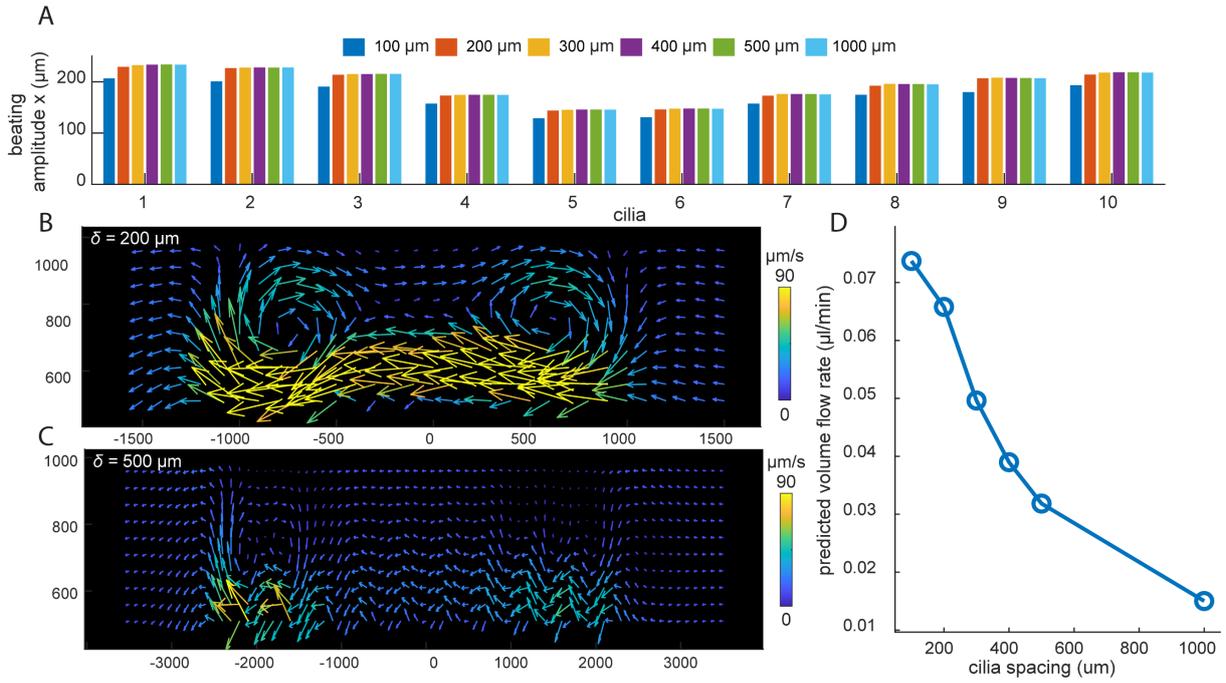

Figure 6: Simulation results of divergent cilia arrays with different cilia spacings at 2 Hz beating frequency. (A) The beating amplitude in the x direction for every cilia. It can be seen that the amplitude is stable above 200 μm cilia spacing. (B) The flow structure with 200 μm cilia spacing, generated by the particle tracking postprocessing of the flow, and it shows a similar vortex structure as in the experiment (Fig. 4(G)). (C) The flow structure with 500 μm cilia spacing. No vortex structure is observed, and the velocities are smaller than with 200 μm spacing. (D) The predicted volume flow rate for different cilia spacings based on simulation results; the calculation can be found in SI5.

# SI1: Femtosecond Laser Assisted Etching (FLAE) process

## Design and generation of the laser toolpath

To design the laser toolpath, 3D models of the cilia and the channel structure are first made with CAD software. Then the geometry is exported into Alphacam to create surfaces and lines on which the focal point of the laser needs to scan through. The toolpath also takes into account the size of the laser voxel. For the cilia mold, the entire volume of the machined part is scanned with overlapping voxels. For the microfluidic channel, hatch lines and planes are added to the geometry, effectively cutting the channel into smaller blocks to reduce the etching time.

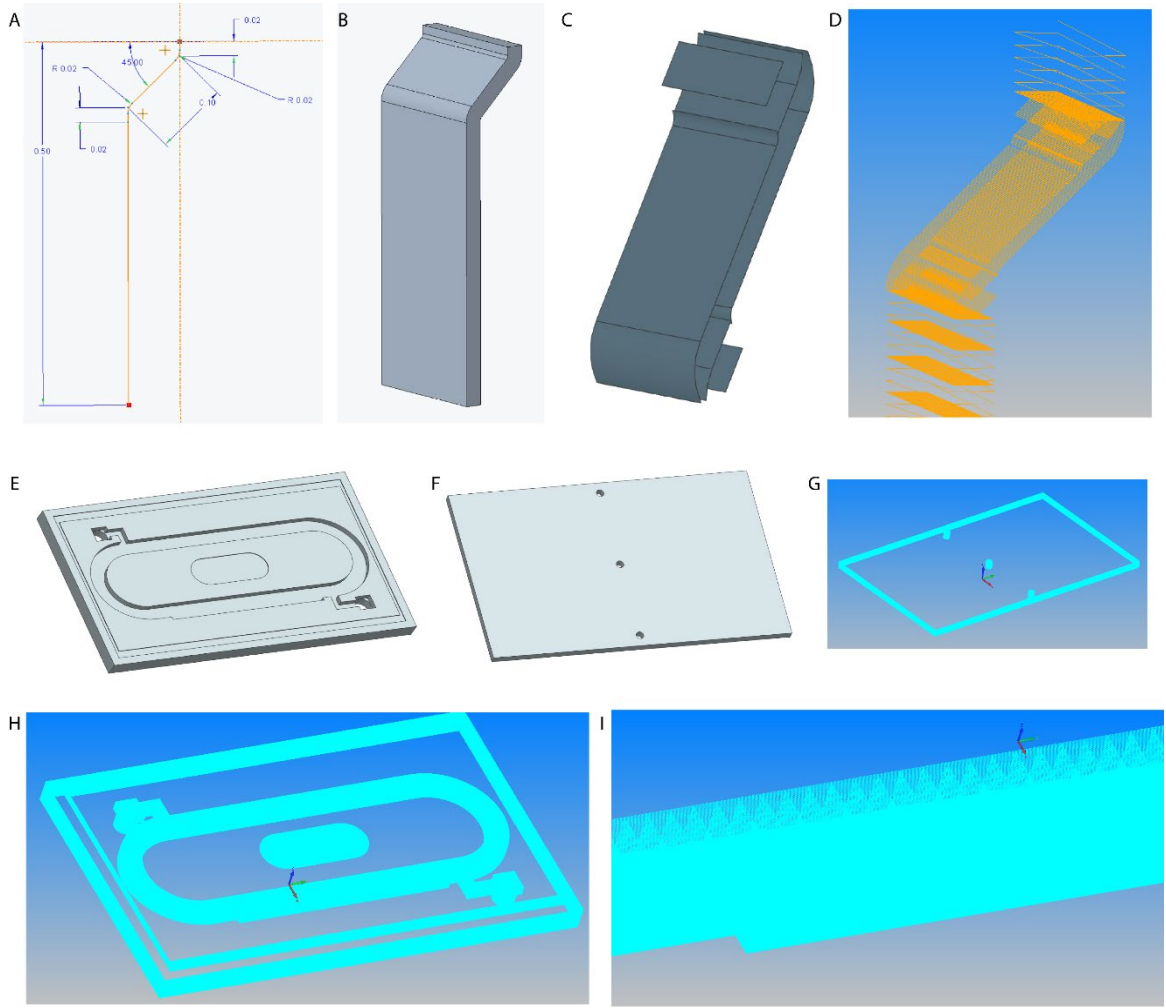

Figure S-1: Geometry design and laser toolpath generation. (A) The center line of the cilium looking from the side. (B) CAD model of the cilia body generated using the swept function in Siemens NX with a 24X200 μm rectangle through the center line. (C) The laser scanning surfaces generated based on the CAD model. (D) The generated laser tool path using Alphacam based on the scanning surfaces. (E) The design of the channel. (F) The design of the channel cap. (G) The laser tool path of the channel cap. (H) The laser tool path of the channel. (I) A closeup shot of the channel laser toolpath showing the hatch lines.

## Fabrication procedure

A femtosecond laser machining system is used to fabricate the cilia molds and the microfluidic channels in fused silica. After laser machining, the slides are placed in potassium hydroxide solution (KOH) bath with ultrasonication to remove the machined parts. The accuracy of the laser machining process is about 1 μm.

For the etching process, the laser machined slide is first submerged a in a container filled with 45 wt% KOH, which is then placed in 85 °C ultrasound bath with the ultrasound intermittently turning on for one minute every ten minutes. The etching speed is 130 μm/h for laser-machined part and 1μm/h for unmachined part. Therefore, high aspect ratio structures will typically have a tapered edge with a 0.5° taper angle, unless the laser path is designed to compensate that. The total etching time is around 7 hours for cilia molds, 5 hours for the channels, and 3 hours for the channel caps.

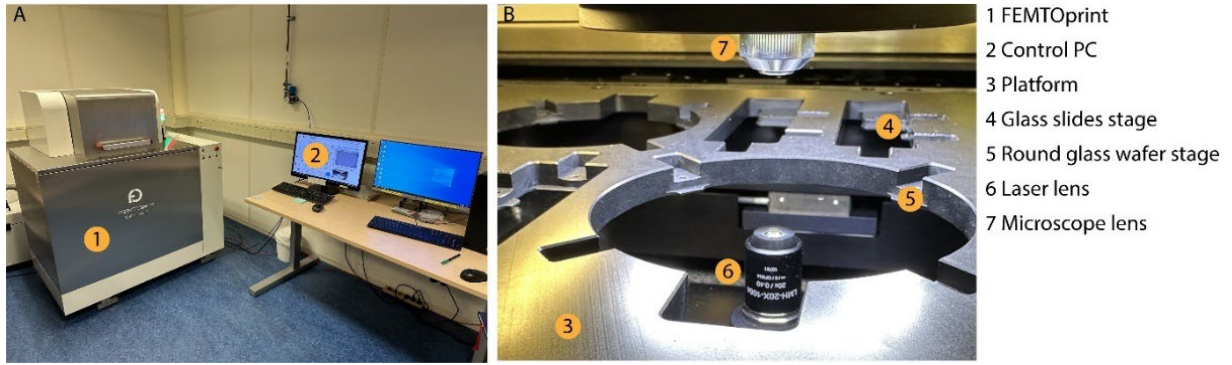

*Figure S-2: FEMTOprint f200 aHead machine. (A) The machine and its control PC. (B) The workpiece holder platform. Glass slides and wafers can be mounted on the platform. The laser lens focus a beam of laser upwards. A microscope placed above the workpiece can be used to monitor the laser machining process.*

# SI2: Cilia fabrication process

## Mixing of magnetic particles and SIBS

The pure SIBS is mixed with magnetic powder using an extruder-mixer as shown in Figure S-3. The magnetic powder (carbonyl-iron powder, ≥99.5%, particle size 5-9 µm, Sigma-Aldrich) account for 70 % of the weight of the final mixture. During mixing, the temperature is set to 150 °C, the counter rotating screws have the same rotation speed of 100 rpm, and the materials are mixed for 10 minutes before releasing.

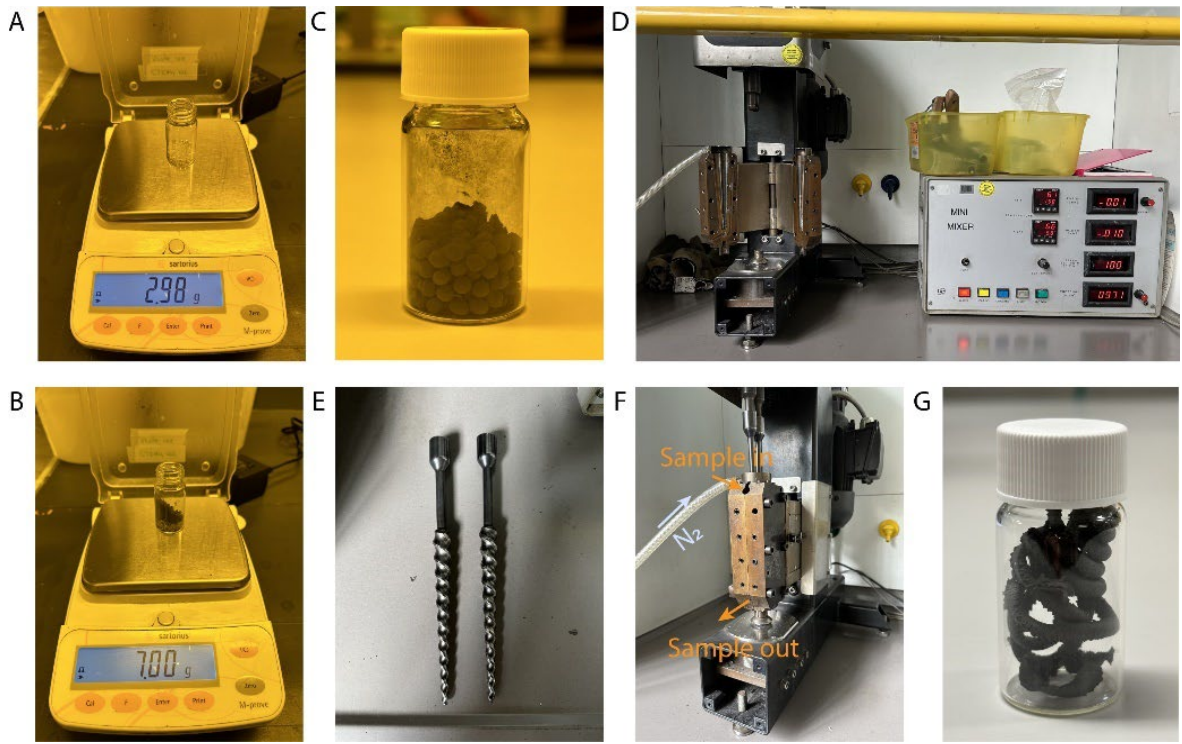

*Figure S-3: A typical process of creating magnetic SIBS. (A) 3 gram pure SIBS. (B) 7 gram carbonyl-iron powder. (C) Pre-mixing of the SIBS and magnetic powder. (D) The twin-screw extruder-mixer. (E) The screws used for mixing. (F) Picture of the extruder-mixer during mixing. The sample is fed in from the upper inlet and discharged from the lower outlet after it is fully mixed. Nitrogen is continuously flowing into the mixing chamber at a low flow rate. (G) The mixed magnetic SIBS.*

## Transfer molding

Molding is performed using a hydraulic press fitted with hot plates (Specac, see Figure S-4). Custom made molds made of brass are used as inserts between the plates for making the thin film precursor layers (Figure S-5) for the cilia fabrication. These molds consist of a round flat plate with milled grooves of varying depths. In our study, we have two molds: one for producing pure SIBS films, with grooves having higher depths, and another for magnetic SIBS films, with grooves of lower depths.

The steps involved in creating pure and magnetic SIBS films are identical. First, raw SIBS pellets were placed above the grooves of the brass mold. Next, a round flat brass plate was used to cover it, and the mold is placed into the press, which is then heated to 180 °C. After allowing the material to soften for approximately 3 minutes, a force of 40 kN was applied, and the temperature was then gradually lowered to 45 °C in approximately 45 minutes. Afterwards, the pressure is released and the resulting thin film can be peeled away from the molds and cut into desired shapes with a blade.

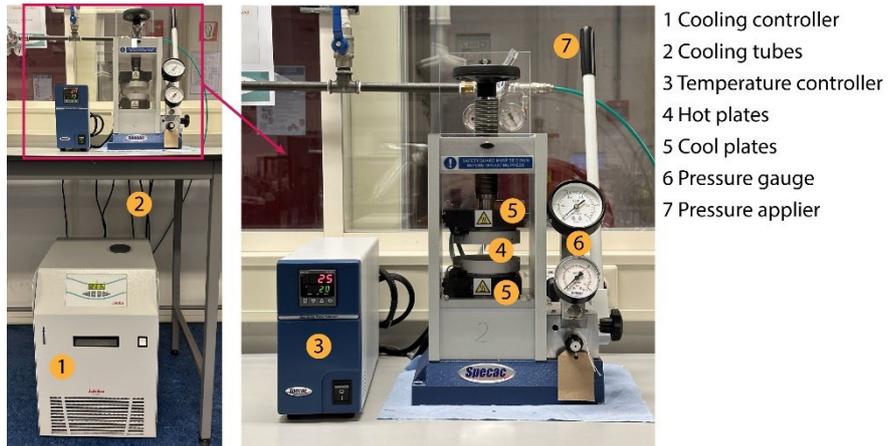

*Figure S-4: The hot press used for transfer molding. The machine contains a cooling system, heating system, and a manual hydraulic press.*

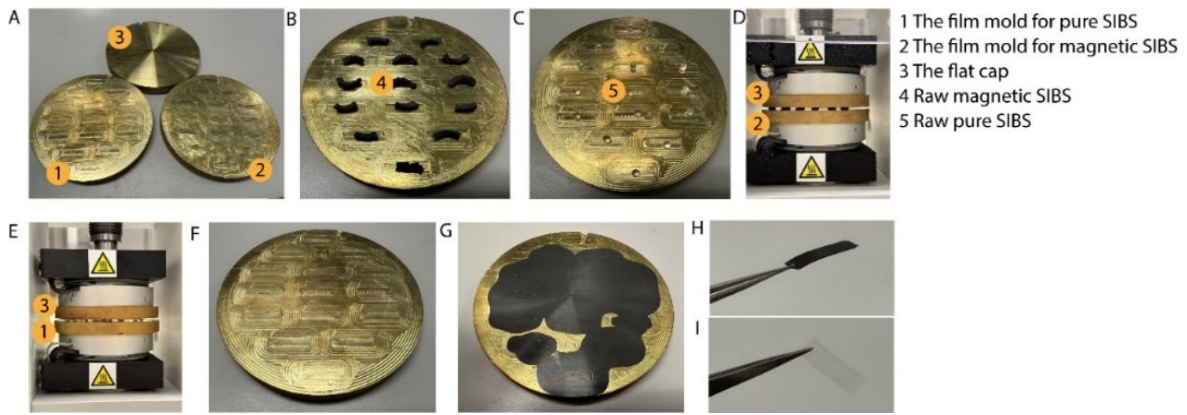

*Figure S-5: The precursor film fabrication process. (A) The brass molds for making thin films. (B) The raw magnetic SIBS placed in the grooves. (C) The raw pure SIBS placed in the grooves. (D) The molding process of magnetic SIBS using the hot embossing. (E) The molding process of pure SIBS. (F)The molded pure SIBS. (G) The molded magnetic SIBS. (H) Magnetic SIBS film cut and demolded. (I) Pure SIBS film cut and demolded.*

After the films are made, they are molded again using the cilia mold made with FLAE and the same hot press. The process is shown in Figure S-6.

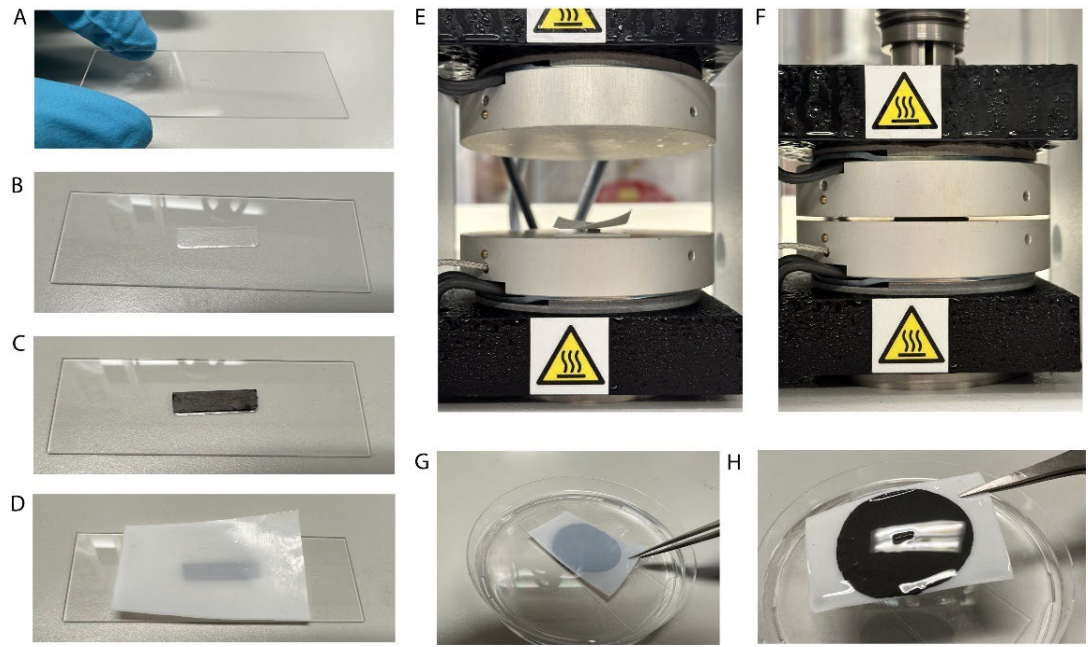

*Figure S-6: Cilia fabrication process. (A) The cilia mold after etching and silanization. (B) Placing the pure SIBS film on the mold. (C) Placing the magnetic SIBS film on the pure SIBS film. (D) Placing a 0.25 mm thick PTFE sheet on the mold. (E) Transferring the materials to the hot press and set the temperature to 130 °C. (F) Applying 4 kN force to the materials for 1 minutes and then lower the temperature to 90 °C. (G) Submerging the mold in isopropanol and demolding. (H) The demolded cilia.*

# SI3: Cilia chip integration and fluid filling

## Cilia integration

After the cilia patches were demolded, they were cut into 0.4 mm wide and 2 mm long stripes and placed sideways into the designated area of the channel. A small drop of low viscosity epoxy glue (Araldite 2020) was added to the gap between cilia patch and the glass surface, and the surface tension helped to spread the glue and flatten the cilia patch to align with the channel bottom (perpendicular to the chip). The glue then is allowed to completely cure for 6 hours at room temperature or 2 hours at 65 °C until the hardening of the glue.

For sealing the channel, a cap with glue filling holes (made with FLAE) was bonded with the cilia-integrated channel. Small amount of the same epoxy glue mentioned above was placed on the filling holes and allowed to spread between the gaps. The surface tension ensures that only the gaps are filled with glue and not the channels.

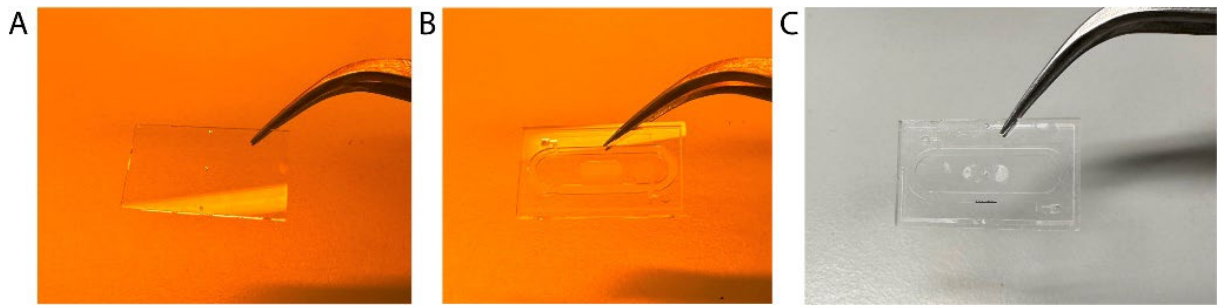

*Figure S-7: The cilia integration process. (A) The channel cap. (B) The channel. (C) The integrated chip with the cilia patch seen at the lower part. First two images were taken inside a cleanroom.*

## Preparation of the tracer particle dispersion and filling of the chip

To achieve neutral buoyancy of the tracer particles (Polystyrene, 5 μm, carboxyl surface, 10 wt%, microParticles GmbH), 20 wt% glycerol solution in DI water was prepared as the carrier fluid. The resulting density is 1059 kg/m$^3$, and the dynamic viscosity is $1.98 \times 10^{-3}$ Pa·s.

30 μl particle suspension was added to 30 ml carrier fluid, resulting in a concentration of 0.01 wt%. The density-matched mixture was then placed in an ultrasonic bath under degas mode for 10 minutes. The mixture is stable without sedimentation for at least 2 months at room temperature.

To fill the chip with the tracer particle suspension, it is first filled with isopropanol using a pipette, as the low surface energy of isopropanol enables it to spread into the gap between cilia without entrapment of air bubbles. DI water was then injected to replace isopropanol. And lastly, the particle solution was injected and the two openings were sealed with silicone adhesive tapes.

For cleaning and storing the chip after the experiment, the tapes were removed and the suspension was washed out first with isopropanol and then with DI water. The chip was then submerged in DI water for long term storage to prevent the cilia from collapsing and irreversibly bonding with each other.

# SI4: Magnetic actuator and recording parameters

## Magnetic actuator

A magnetic actuator was built for actuating the cilia at high frequencies. To achieve sufficient and stable magnetic field at all frequencies, a permanent magnet Halbach array was used instead of an electromagnet. 16 neodymium magnets (10×10×40 mm) were used to generate a uniform field of 0.22 Tesla in the center, The array is driven by a DC motor with an encoder through a timing belt connection to allow light passing through the optical path. The chip is then placed in the middle plane of the array, where the cilia patch is situated at the center of the circle to ensure maximum homogeneity of the field.

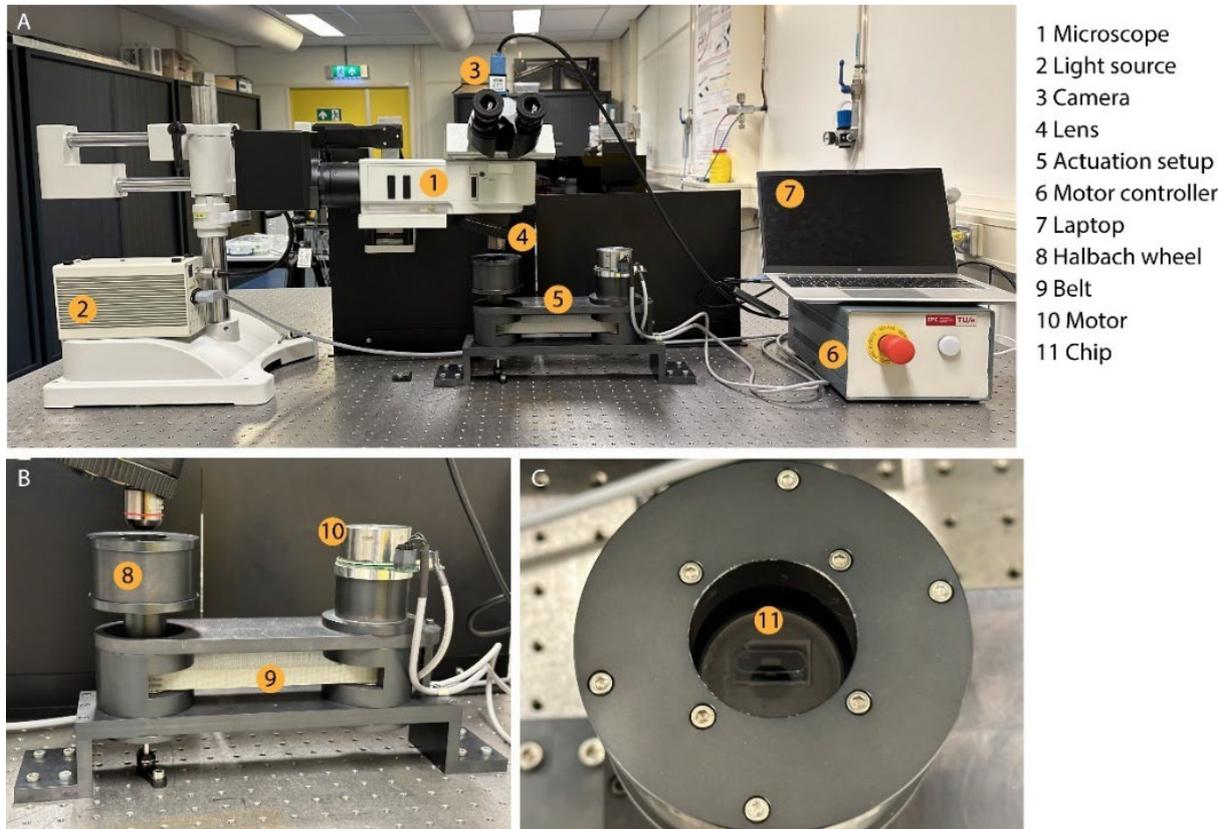

*Figure S-8: Experiment setup. (A) The entire experiment setup consists of a microscope, the magnetic actuator, and a computer. (B) The magnetic actuator consists of a DC motor, a Halbach wheel and other mechanical components for the connection. (C) The chip placed on the chip stage holder in the central area of the Halbach wheel.*

## Recording setup

The raw data collected in this study is mainly videos captured by cameras. The detailed information of the data collection is shown in the table below.

Table 1 Experimental and recording parameters

| Measurement objective | Lens | Camera | ROI | Rotation frequency (Hz) | Cilia beating / actuation frequency (Hz) | Video capture speed (fps) | Exposure time (sec) |
|---|---|---|---|---|---|---|---|
| Fully developed flow | Olympus LMPLFLN 5XBD | IMAGINGSOURCE DMK 33UX252 | 1 | 1 | 2 | 1 | 1/1000 |
|  |  |  |  | 10 | 20 | 10 |  |
|  |  |  |  | 20 | 40 | 20 |  |
|  |  |  |  | 30 | 60 | 30 |  |
|  |  |  |  | 40 | 80 | 40 |  |
|  |  |  |  | 50 | 100 | 50 |  |
| Instantaneous flow field |  |  | 2 | 1 | 2 | 180 |  |
| Long-term particle trajectory |  |  |  | 1 | 2 | 2 |  |
|  |  |  |  | 10 | 20 | 20 |  |
| | | | | 20 | 40 | 40 | |
| | | | | 30 | 60 | 60 | |
| | | | | 40 | 80 | 80 | |
| | | | | 50 | 100 | 100 | |
| Cilia motion | Olympus LMPLFLN 20XBD | Phantom VEO 1310 high-speed camera | | 1 | 2 | 100 | 1/42000 |
| | | | | 10 | 20 | 1000 | |
| | | | | 20 | 40 | 2000 | |
| | | | | 30 | 60 | 3000 | |
| | | | | 40 | 80 | 4000 | |
| | | | | 50 | 100 | 5000 | |

## SI5: Net flow velocity prediction from COMSOL simulation

In the 2D simulation, we have kept the channel height and the cilia geometry the same as in the experiments. Before calculating the net flow, the cilia dynamic and the local vortices were first compared with experimental results for validation, as shown in the main text with Figure 2 and 3. Due to the laminar nature of the fully developed channel flow away from the cilia array, COMSOL simulations were set up without simulating the whole channel length as in the experiments, but only a sufficiently long, open ended channel that the cilia driven flow is allowed to develop into a complete parabolic profile. In this way, we ensure the reliability of the results, while saving the computation cost.

To translate the simulated flow rate to the actual predicted flow rate in the channel, a simple conversion is performed. The channel length in the simulation is 10 mm, with the cilia array occupying the middle 2 mm, so the effective length of the channel is about 8 mm. In the experiment, the total length of the recirculating channel is 36 mm excluding the cilia patch, hence 4.5 times the simulated channel length. Due to the laminar nature of the flow, it is then reasonable to assume that the flow resistance is proportional to the channel length, therefore the estimated flow rate is calculated by dividing the simulated flow rate by 4.5. The resulting volume flow rate at actuation frequency of 2 Hz is about 0.066 µl/min, compared to 0.039 µl/min in the experiments. Considering the difference in 2D (single parabolic profile) and 3D (semi-double parabolic profile), we can say that the simulation result matches reasonably well with the experiments.

**Movie 1**  01_highspeed_tracking.mp4

Higher time-resolution imaging showing time-dependent flow structures with divergent and convergent arrays actuated at 2 Hz. The particles perform cyclic motion with net displacement over time. Walking vortices can be observed during each beating cycle of the cilia.

**Movie 2**  02_strobo_convergent.mp4

Stroboscopic imaging showing particle trajectories over longer durations with the convergent array. The video capture rate is the same as the actuation rate of cilia, so the differences in consecutive frames show the net displacement of particles. Stable vortices can be observed from these trajectories.

**Movie 3**  03_ strobo_divergent.mp4

Same as Movie 2, but with the divergent array.

**Movie 4**  04_simulation.mp4

Simulation results showing the time-dependent streamlines and velocity map.

**Movie 5** 05_net_flow.mp4

Particle tracking at ROI1 location on the chip with the divergent cilia array actuated at 2 Hz.